\documentstyle[12pt]{article}

\catcode`\@=11

\newcommand{\preprintno}{preprint number here}	 

\def\@maketitle{\newpage   
 \null
 \vspace*{-1\headsep}      
 \vspace*{-1\headheight}
 \vspace*{-24pt}
 \begin{flushright}{\absize
   { \preprintno} }
 \end{flushright}
 \vskip \headsep	   
 \vskip \headheight
 \begin{center}		   
   {\tisize\bf \@title \par}
   \vskip 2em
   {\ausize
     \begin{tabular}[t]{c}\@author
     \end{tabular}\par}
   \vskip 1ex
 \end{center}
 \par
 \vskip 2ex}

\def\abstract{
\absize
\begin{center}
{\bf Abstract\vspace{0pt}}
\end{center}
\setskip
\quotation
}

\def\appendix{\par				
    \setcounter{section}{0}			
    \setcounter{subsection}{0}
    \renewcommand{\theequation}{\Alph{section}.\arabic{equation}}
    \renewcommand{\thesection}{Appendix \Alph{section}
		\setcounter{equation}{0}  } 
}

\def\section{
\setcounter{equation}{0} 	
\@startsection {section}{1}{\z@}{-3.5ex plus -1ex minus
 -.2ex}{2.3ex plus .2ex}{\Large\bf}}

\renewcommand{\theequation}{\arabic{section}.\arabic{equation}}

\def\@versim#1#2{\lower 2\p@\vbox{\baselineskip\z@skip\lineskip+1\p@
    \ialign{$\m@th#1\hfil##\hfil$\crcr#2\crcr\sim\crcr}}}
\def\gapp{\mathrel{\mathpalette\@versim>}}
\def\lapp{\mathrel{\mathpalette\@versim<}}

\catcode`\@=12

\let\al=\alpha
\let\be=\beta

\let\Ga=\Gamma
\let\de=\delta
\let\De=\Delta
\let\ep=\varepsilon

\let\la=\lambda
\let\La=\Lambda

\let\Si=\Sigma

\let\p=\partial
\let\<=\langle
\let\>=\rangle

\let\txt=\textstyle
\let\dsp=\displaystyle

\def\eqn#1{(\ref{#1})}      
\def\Eqn#1{Eq.~(\ref{#1})}  
\def\e{{\rm e}}
\def\beq{\begin{equation}}
\def\eeq{\end{equation}}
\def\ba{\begin{array}}
\def\bea{\begin{eqnarray}}
\def\ea{\end{array}}
\def\eea{\end{eqnarray}}

\def\comment#1{ \hbox{[{\it Comment suppressed here.}\/]} }
\def\hide#1{}
\def\tr{\hbox{tr}}

\def\kv{ {\bf k} }
\newdimen\pmboffset
\def\oldpmb#1{\setbox0=\hbox{#1}%
 \copy0\kern-\wd0
 \kern\pmboffset\raise 1.732\pmboffset\copy0\kern-\wd0
 \kern\pmboffset\box0}

\def\tisize{\Large}      
\def\ausize{\normalsize} 
\def\absize{\normalsize} 
\def\setskip{ \setlength{\baselineskip}{3ex} } 

\def\starttext{
\setlength{\baselineskip}{ 17pt} 
\pagenumbering{arabic}
}

\def\half{{\txt{1\over 2}}}

\def\third{{\txt{1\over 3}}}
\def\sixth{{\txt{1\over 6}}}
\def\twelfth{{\txt{1\over 12}}}
\def\twfourth{{\txt{1\over 24}}}
\def\f{\mskip -2mu f}   
\def\O{{\cal O}}	
\def\phb{\overline\phi}
\def\wb{\bar w}
\def\eg{ e.g.~}
\def\ie{ i.e.~}
\def\etal{{\it et al.}}
\def\phib{\overline{\varphi}}
\def\pt{{\rm PT}}
\def\GeV{{\rm GeV}}
\newenvironment{equationwithlabel}[1]
  {
  \begin{equation}\label{#1}}{\end{equation}}

\newcommand{\beql}[1]{\begin{equationwithlabel}{#1}}
\newcommand{\eeql}{\end{equationwithlabel}}

\begin{document}

\title{Radiatively-Induced First-Order Phase Transitions:
The Necessity of the Renormalization Group}

\author{
        Mark Alford\\[0.5ex]
	Laboratory of Nuclear Studies,\\
	Cornell University,\\
	Ithaca, NY 14853\\[2ex]
        and\\[2ex]
	John March-Russell\\[0.5ex]
	Theoretical Physics Group,\\
	Lawrence Berkeley Laboratory,\\
	1 Cyclotron Rd,\\
	Berkeley, CA 94720
}

\renewcommand{\preprintno}{CLNS 93/1244\\ LBL-34573 \\ hep-ph/9308364}

\begin{titlepage}
\maketitle
\def\thepage{}		

\begin{abstract}
We advocate a (Wilson) renormalization-group (RG) treatment
of finite-temperature first-order phase transitions, in
particular those driven by radiative corrections such as
occur in the standard model, and other spontaneously-broken
gauge theories.  We introduce the scale-dependent coarse-grained
free energy $S_\La[\phi]$ which we explicitly calculate, using
the Wilson RG and a $(4-\ep)$-expansion, for a scalar toy
model that shares many features of the gauged case.
As argued by Langer and others, the dynamics
of the phase transition are described by $S_\La[\phi]$ with
$\La$ of order the bubble wall thickness, and {\it not} by
the usual (RG-improved) finite-temperature effective action
which is reproduced by $S_\La[\phi]$ for $\La\to 0$.
We argue that for weakly first-order transitions
(such as that in the standard model) the $(4-\ep)$-expansion
is necessary to control an inevitable growth of the effective
scale-dependent coupling towards the strong-coupling regime,
and that diagrammatic resummation techniques are unlikely
to be appropriate.
\
\end{abstract}

\end{titlepage}

\renewcommand{\thepage}{\roman{page}}
\setcounter{page}{2}
\mbox{ }

\vskip 1in

\begin{center}
{\bf Disclaimer}
\end{center}

\vskip .2in

\begin{scriptsize}
\begin{quotation}
This document was prepared as an account of work sponsored by the United
States Government.  Neither the United States Government nor any agency
thereof, nor The Regents of the University of California, nor any of their
employees, makes any warranty, express or implied, or assumes any legal
liability or responsibility for the accuracy, completeness, or usefulness
of any information, apparatus, product, or process disclosed, or represents
that its use would not infringe privately owned rights.  Reference herein
to any specific commercial products process, or service by its trade name,
trademark, manufacturer, or otherwise, does not necessarily constitute or
imply its endorsement, recommendation, or favoring by the United States
Government or any agency thereof, or The Regents of the University of
California.  The views and opinions of authors expressed herein do not
necessarily state or reflect those of the United States Government or any
agency thereof of The Regents of the University of California and shall
not be used for advertising or product endorsement purposes.
\end{quotation}
\end{scriptsize}

\vskip 2in

\begin{center}
\begin{small}
{\it Lawrence Berkeley Laboratory is an equal opportunity employer.}
\end{small}
\end{center}

\newpage
\renewcommand{\thepage}{\arabic{page}}
\setcounter{page}{1}

\starttext

\section {Introduction}

Systems that undergo first-order phase transitions as a function
of temperature have been extensively discussed
in the particle physics literature during the last decade.
The initial motivation was the role of GUT-scale
first-order phase transitions in driving
an inflationary period of expansion in the very early universe
\cite{inflation}. Lately the primary motivation has been
to show that the observed baryon
asymmetry of the universe can be produced through the
baryon and lepton number anomalies, in concert with
the out-of-equilibrium evolution of a first-order
electroweak phase transition \cite{sphaleron}.

Most of the systems studied in these two contexts have been
gauge theories where the first-order transition
is {\it radiatively induced} ({\it fluctuation-driven} in the
condensed-matter terminology we will employ in this paper).
This means that the tree-level scalar potential
has a form that implies a second-order phase transition,
but when fluctuations in the fields are taken into account (at
one-loop or higher) the potential indicates a first-order phase
transition. If the transition is strongly first-order then
the effective potential may be calculated by the standard
loop expansion. However if it is weakly first-order
(generically, when $(M_h/M_v)^2 \gapp 1$, where
$M_h$ and $M_v$ are the Higgs and gauge boson masses)
then the naive loop expansion breaks down,
we must use the renormalization group (RG) to obtain reliable
results \cite{HLM}\cite{jmr}.
Ginsparg \cite{Ginsparg} has investigated the Gell-Mann--Low
RG flows of Yang-Mills-Higgs systems in the heavy Higgs regime,
and found that the phase transition remains first order
(specifically, there were no stable fixed points within the
$(4-\ep)$-expansion) for reasonable scalar sectors.

In this paper we propose to study the properties of
weakly first-order phase transitions
by the systematic use of the Wilson RG,
together with an expansion in the number of spatial dimensions
around four (the $(4-\ep)$-expansion) \cite{Wilson}.
Following earlier work in the condensed matter
literature, we will take the view that the appropriate quantity
for the analysis of a first-order phase transition is the {\it scale
dependent} (``coarse-grained'')
effective Hamiltonian defined by successively
integrating out momentum modes. In the limit of infinite
coarse-graining scale, this coincides with the effective action,
and correctly describes the static properties of the system.
However for a correct description of the dynamics we must
coarse-grain only up to a physical scale, typically the correlation
length. This approach has been advocated
by Langer and coworkers \cite{Langer}, and by
Kawasaki, Imaeda, and Gunton  \cite{KIG}, and is also, of course,
the strategy proposed by Wilson in his work on critical phenomena.
We will explicitly calculate the coarse-grained effective potential
for a toy model which possesses a fluctuation-induced first-order phase
transition similar to that of the standard model.

An important motivation for our RG study is that the current experimental
lower bound on the Higgs mass is around 60 \GeV,
so that the naive loop expansion for the electroweak phase transition
is breaking down.
Over the last few years there have been many attempts in
the particle physics literature to remedy this situation
\cite{sphaleron}. They have mainly consisted of
attempts to sum certain infinite subclasses of diagrams
(for example ``daisy'' or ``super-daisy''
graphs, and ``ring-improvement'') and thus improve perturbation theory.
We believe that the RG analysis is an advance over these
diagrammatic methods for three reasons.

Firstly, the RG analysis that we will discuss
below generically leads to results for the properties of the phase
transition (such as latent heats, bubble wall tensions etc.)
that are highly non-analytic as a function of initial couplings (and
$\ep$). This non-analytic behaviour is quite easily handled
by the RG, but is very difficult to reproduce by diagrammatic
methods.

Second, as we will see in detail below, the typical flows of the
effective couplings are such as to take them first towards, and
then away from, an unstable fixed point.
These unstable fixed points occur at values of the couplings of
order $2\pi^2\ep$. As $\ep\to 1$ (the physical value) this
is far into the region of strong coupling where
even the (non-Borel-resummed)
renormalization group improved loop expansion
is invalid. Diagrammatic methods obscure this
problem rather than solving it. In this paper we will
take $\ep \ll 1$ for most of our analysis, only setting
$\ep=1$ at the final stage. The motivation, and dangers \cite{jmr},
of this approach we will discuss at the end of the Introduction,
and in Section~5.

Thirdly, the Wilson RG flow of a theory down to a scale $\La$ is
{\it by definition} the procedure which gives us
the correct Hamiltonian for the dynamics of the modes of
momentum $|k|<\La$, when fluctuations on scales $|k|>\La$ are
taken into account (integrated out). Thus, at a sensibly
chosen scale, it is the relevant
object for studying the dynamics of the phase transition.
For example, the properties of the critical bubbles that nucleate
the phase transition, such as the bubble wall tension, are
determined by the effective Hamiltonian with coarse-graining
length of order the correlation length \cite{Langer}.
We show that the effective Hamiltonian at this coarse-graining
scale {\it calculably differs} from the usual effective action
(which corresponds to an infinite coarse-graining length).

Rather than confront the full complication
of the phase transition of a GUT or the electroweak model,
in this paper we will study the following purely scalar system:
\beql{I:model}
{\cal L} = {1\over 2} (\p_\mu \phib)\cdot(\p^\mu \phib)
          - {\mu^2\over 2} \phib\cdot\phib
          +  {h_1\over 4!} (\phib\cdot\phib)^2
               + {h_2\over 4!} \sum_{i=1}^{N} (\phib_i)^4.
\eeql
Here $\phib$ is an $N$-component real scalar field, and $\phib\cdot\phib
=\sum_{i=1}^{N} \phib_i\phib_i$. Inclusion of the second quartic
coupling $h_2$ breaks the original continuous $O(N)$ symmetry of
the model down to the discrete hypercubic group in $N$-dimensions.
This system exhibits a close analog to the weakly-first-order
fluctuation driven phase transition that occurs in
the standard model (and most other Yang-Mills-Higgs systems)
for suitable Higgs mass. This will
enable us to discuss some very important qualitative
features of the physics of such phase transitions
that have been missed by previous authors,
without the inessential complication of gauge invariance.
In the final section of this paper we will explain why our toy system
is indeed a good theoretical laboratory for the general, gauged, case.

There is one caveat that we should mention at this point.
As one of has argued in an earlier
publication \cite{jmr} the results of a $4-\ep$ expansion can be
unreliable; in particular, new fixed points of the RG can appear
at $\ep=1$ that are not accessible by such an expansion,
rendering the transition second-order after all.
However very useful information
can still be extracted {\it if} the phase transition really is
first-order, since its properties (such as the latent heat, critical
bubble parameters {\it etc.}) can be calculated as an (asymptotic)
expansion in $\ep$.
Of course, the leading order calculation that we undertake
in this work is not sufficient to accurately determine the
properties at $\ep=1$. However we are hopeful that higher-order calculations,
in concert with asymptotic estimates of large-order $\ep$-perturbation
theory, can supply accurate parameters in $D=3$, as is the case
for second-order phase transitions \cite{Zinn-Justin}.
Whether or not this succeeds, we have still extracted information about
the limit $\ep\to 0$ which other proposed calculations must
be able to reproduce if they are correct.

Before starting, we note that
the RG has been previously applied to the cubic anisotropic
model \eqn{I:model} by Rudnick \cite{Rud}, however he specialized to
$N=2$, used classical statistical mechanics rather than finite
temperature field theory,
and used specific methods whose generalization to more
complicated (\eg~gauge) theories is not immediate; and also
by Amit \cite{Amit}, who used the Gell-Mann--Low RG in which there is
no finite coarse-graining scale.
We have derived our evolution equations from
Polchinski's general formulation \cite{Pol},
using the perturbative expansion of Bonini, D'Attanasio,
and Marchesini \cite{BDM}. This makes it relatively straightforward
to push our results to higher order in the couplings, and
even to include the evolution of derivative terms.

Finally, we should mention that the philosophy of coarse-graining has
also been advocated recently by Tetradis and Wetterich \cite{TW}, whose
``average potential'' is a smoothly cut-off version of the Wilson
coarse-grained free energy (however, for a critique of the smooth cutoff
see Ref.~\cite{Mor}). They obtain accurate critical exponents for
the second-order phase transition in an $O(N)$ model, but do
not study first-order phase transitions.

\section{The coarse-grained potential}

The Wilson RG \cite{Wilson} is based on the idea of gradually
integrating out degrees of freedom in a field theory by lowering
the ultraviolet cutoff $\La$,
and at the same time varying the action functional in such
a way that all Green functions of the remaining modes are
invariant. We first describe how this is accomplished for
zero temperature Euclidean quantum field theories, or
equivalently classical statistical mechanical systems,
and then generalize to finite temperature quantum field theories.

\subsection{The coarse-grained potential at zero temperature}

An elegant account of Wilson coarse-graining was given by
Polchinski \cite{Pol} (see also Morris \cite{Mor}).
Following his analysis we consider a simple
scalar field theory defined by the functional integral
\beql{II:gfunct}
\ba{rcl}
Z[J] &=& \dsp \int [D\phi]
              \exp\biggl\{ \int_k \Bigl[-\half\phi(k)(k^2+m^2)
              K^{-1}(k^2/\La^2)\phi(-k) + J(k)\phi(-k)\Bigr] + L_\La[\phi]
              \biggr\}\\
        &\equiv & \dsp \int [d\phi] \exp S_\La[\phi] ,
\ea
\eeql
where $K(k^2/\La^2)$ is a smooth UV cutoff function that falls
quickly to zero for $k^2 > \La^2$. The generating functional, $Z$, is
independent of $\La$
(up to physically irrelevant field-independent changes
of overall normalization) if the interaction Lagrangian $L$ satisfies
\beql{II:poleqn}
\dsp \La { dL[\phi]\over d\La}
      = \dsp -\half \int_k {(2\pi)^8 \over
         (k^2 + m^2)} \La {dK(k^2/\La^2)\over d\La}
      \dsp\biggl\{ {\de L\over \de\phi(k)}{\de L\over \de\phi(-k)}
         + {\de^2 L\over \de\phi(k)\de\phi(-k)} \biggr\}
\eeql
together with $J(k)=0$ for $k^2 > \La^2$. The Green's functions of all
the modes $\phi(k)$ with $k^2 \leq \La^2$ are then independent of $\La$.
No subtleties arise in the
verification of these formulae since, with a physical cutoff,
no divergences are present.
The resultant coarse-grained effective action
$S_\La[\phi]$ is the action that describes the dynamics of the
degrees of freedom with momentum $k \leq \La$. In the infinitely
coarse-grained limit $\La \to 0$ it becomes the standard effective
action.

Note that the evolution equation \eqn{II:poleqn} is an {\it exact} functional
equation for the coarse-grained effective action. We wish
to use this equation (actually a generalization to finite temperature)
to study the RG flow of our model system \Eqn{I:model}. It is, however, in
an inconvenient form for our purposes -- in particular the derivation
of the ``beta-functions'' for the flow of the couplings is rather involved
in this formulation.

We will therefore employ the Wilson RG in a formulation
originally constructed by Wegner and Houghton
\cite{WH} in a statistical mechanical context,
and further elaborated by Nicoll, Chang, and Stanley \cite{NCS}.
This formulation was then used to give the first logically
complete derivation of a coarse-grained effective action by
Kawasaki, Imaeda, and Gunton (KIG) \cite{KIG}. (There are
other discussions of this topic in the condensed-matter
literature, see \cite{GD} and \cite{Binder}, but
overall there has been very little work on the
derivation of the coarse-grained free-energy from
a microscopic Hamiltonian.)
Our choice of toy system \Eqn{I:model} allows one major technical
simplification in the passage from the Wilson-Polchinski equation
\eqn{II:poleqn} to the KIG form. Since our model involves only
scalar fields with quartic interaction terms in the initial
action, wavefunction renormalization only arises at two loops
and above in the perturbative (coupling constant) expansion.
This in turn implies that wavefunction
renormalization is an $\O(\ep^2)$ effect. However we will see that
a calculation to $\O(\ep)$ is sufficient to uncover the properties
of the phase transition, and therefore, we can neglect the
renormalization of the kinetic term to the order in which we work.
Furthermore, all higher derivative terms generated as a result
of the RG procedure are similarly higher order in the $\ep$-expansion.
(Of course at very large momentum -- in other words for field
configurations with large spatial derivatives -- the higher
derivative terms can dominate the kinetic term. However for
the configurations of interest, such as true vacuum bubbles, the
gradients are such that this estimate is correct.)
We can therefore ignore the renormalization of {\it all}
momentum dependent terms -- only the potential terms
are renormalized to $\O(\ep)$. This is a significant
technical simplification relative to, say, scalar QED, or
any other gauged system, where we would have to include
wavefunction renormalization. We note in passing that it is possible
to generalize our analysis to this case -- this work
will be reported in a future publication.

In the case where we can ignore the flow of the derivative
terms, the Wilson-Polchinski equation \eqn{II:poleqn}
can be recast in the form:
\beql{II:Kaw}
{\p V_\La(\phib) \over \p\La} = {K_{(D+1)}\over 2} \La^D
   \tr_{ij}\ln\biggl[ \La^2\de_{i,j} +
     {\p^2 V_\La \over \p\phib_i \p\phib_j} \biggr]~.
\eeql
This is our master equation for the flow of the potential of an
$N$-component scalar field theory in $(D+1)$ Euclidean dimensions.
Here $K_{(D+1)}=S_{D}/(2\pi)^{(D+1)}$, and $S_{D}$ is the surface
area of a unit $D$-sphere.
This equation was first explicitly written down in Ref.~\cite{KIG},
so we will refer to it as the KIG-equation.
In Appendix A we show that it can be derived from
Ref's.~\cite{BDM} perturbative analysis of Polchinski's exact RG flow
equation. Intuitively, however, \Eqn{II:Kaw} seems very
reasonable: it states that the contribution to the effective
potential from integrating out the degrees of freedom in an
infinitesimal shell $\La-\de\La < |k| < \La$ in momentum space is just
the standard loop result \cite{Sid}
for the effective potential, with the momentum integral
restricted to that shell.

The KIG equation as formulated above describes the flow of
statistical
mechanical systems, which correspond to zero temperature
Euclidean quantum field theories.
Given the potential at some high cutoff $\La_0$, it shows us
how to flow down to some scale of interest $\La_1$ and
obtain $V_{\La_1}(\phi)$. Following KIG, we proceed in two stages:
\begin{enumerate}
\item Expand $V_\La(\phi)$ as a truncated power series in $\phi$,
including all the relevant operators (here defined to include the
marginal operators as well). Recall that relevant operators
are those whose couplings do not have large negative mass dimension
for small $\ep$.
We obtain coupled differential equations
for the relevant couplings (only), which in the case of $\phi^4$ theory
near four space-time dimensions
are the mass-squared $m^2$ and the $\phi^4$ coupling(s) $h_i$,
which we solve to
obtain $m^2(\La)$ and $h_i(\La)$.
\item Substitute the solutions of these equations into the RHS
of the KIG equation, and integrate from $\La_0$ to $\La_1$
to obtain the higher order corrections to $V_{\La_1}$,
\ie $\phi^6$ and higher.
In general one must check for self-consistency---the ``back-reaction''
effect of these truncated couplings on the differential equations
for the relevant couplings must be higher order in some small parameter.
\end{enumerate}
Note the standard 1-loop effective potential (with ultraviolet cutoff
$\La_0$ and infra-red cutoff $\La_1$)
may be obtained by the crude approximation of
skipping the first step, and substituting $m(\La_0)$ and
$g(\La_0)$ directly into the RHS in the second step.

In practical applications the theory would be specified by
renormalization conditions on $m^2$ and the $h_i$ at a low momentum scale,
so that at stage 1 we would choose values of $m^2(\La_0)$ and
$h_i(\La_0)$ that give the right values when flowed down using
\Eqn{II:Kaw}, then at stage 2 we would calculate the remaining
terms in the low-momentum potential. This is done because
experimental measurements are typically made at energy scales
low relative to the UV cutoff. In this paper, however, we
are dealing with a toy model, and hence will express our results
in terms of couplings defined at a high energy scale $\La_t$,
defined in the next subsection.

\subsection{Coarse-grained potential at finite temperature}

The generalization of the KIG flow equation to finite-temperature
becomes obvious if we recall the rules for constructing
the finite-temperature (equilibrium) generating functional for a
field theory: We compactify the Euclidean time direction on a circle of
circumference $\be=1/T$. This results in a discrete set
of fourier modes $\phib_{i,n}(\kv)$, with $n\in Z$ -- the Matsubara modes.
The $(D+1)$-momentum becomes $k=(2\pi n T,\kv)$, and integrals over $k_0$
are replaced by a sum over $n$.
For example, the finite-temperature generating functional for
a simple $\la \phi^4$ theory, is, in momentum space:
\beql{II:genfuncT}
\ba{rcl}
\dsp Z[J,T] =& &\dsp \int [D\phi] \exp\biggl\{ - T \sum_n \int_\kv
           \bigl(-\half(\kv^2+(2\pi nT)^2 - m^2)\phi_n(\kv)
           \phi_{-n}(-\kv) + J\cdot\phi \bigr)\\
           &-& \dsp {\la\over 4!} T^3 \sum_{n,n',n''} \int_{\kv,\kv',\kv''}
           \phi_n(\kv) \phi_{n'}(\kv')\phi_{n''}(\kv'')
           \phi_{-n-n'-n''}(-\kv-\kv'-\kv'')
           \biggr\}.
\ea
\eeql
It will be convenient to simplify this expression so that we have a
conventionally normalized kinetic term -- this is achieved by
rescaling the field $\phi_n(\kv)$ by a factor of $\sqrt{T}$.
With this choice, the quartic term is proportional to $\la T$.
This is convenient since a field theory in three dimensions has
a quartic coupling of mass dimension one, which in this case is
soaked up by $T$, so $\la$ stays dimensionless.

In any case, the important point about \Eqn{II:genfuncT} is that
it is of the form of (infinitely) many coupled ``species''
of scalar field, indexed by $n$, living in one lower
dimension. Therefore we can directly apply the KIG equation
\eqn{II:Kaw} to this system -- actually the $N$-component version
-- to find,
\beql{II:Obvious}
{\p V_\La^T(\phib) \over \p\La} = {K_{D}\over 2} \La^{(D-1)}
   \tr_{(i,j), (n,m)}\ln\biggl[ \La^2\de_{i,j}\de_{n,-m} +
     {\p^2 V_\La^T \over \p\phib_{i,n} \p\phib_{j,m} } \biggr]~,
\eeql
as the appropriate generalization of the KIG equation to
finite-temperature. Remember that in this equation the
potential now includes a temperature-dependent effective
$({\rm mass})^2$ term, $(2\pi nT)^2$, for the $n\not=0$ modes.
(We employ the $\tr_{(n,m)}$ notation as a shorthand for a sum over
Matsubara modes where the discretized energy is conserved at each vertex.)

There are two interesting limits of this formula.
For $\La\gg 2\pi T$, the Matsubara frequencies are so close that the
``trace'' over $n$ approximates the original $(D+1)$-dimensional integral,
so \Eqn{II:Kaw} is valid.
For $\La \ll 2\pi T$, the sum is dominated by the $n=0$ mode --
the very massive $n\not=0$ modes decouple leaving
behind temperature-dependent corrections to the potential, and
the system looks $D$-dimensional.
In this limit \Eqn{II:Obvious} thus simplifies to
\beql{II:Hitemp}
{\p V_\La^T(\phib_0) \over \p\La} = {K_{D}\over 2} \La^{(D-1)}
   \tr_{i,j}\ln\biggl[ \La^2\de_{i,j} +
     {\p^2 V_\La^T \over \p\phib_{i,0} \p\phib_{j,0} } \biggr]~,
\eeql
with the addition of a particular, temperature dependent,
$V^T_\La(\phib_0)$ as a
boundary condition at the starting scale $\La_t < T$.

We must remember, however, that for a finite temperature field theory
the renormalization conditions are still imposed on the zero
temperature theory. The correct procedure is therefore
to select some high scale $\La_0 \gg T$, and choose $m^2(\La_0)$
and $g(\La_0)$ such that the renormalization conditions are obeyed
when we flow down to low momentum in the zero temperature theory,
{\it i.e.}~using \eqn{II:Kaw}. To calculate the finite temperature
coarse-grained effective potential we should
in principle run down from $\La_0$ using \Eqn{II:Obvious}, however
the sum over Matsubara frequencies is difficult to perform
analytically.

Therefore in this paper we will instead arbitrarily choose some
scale $\La_t$ somewhat below $T$, so that we can
make the reasonable approximation of completely integrating out
the massive $n\not=0$ modes. The running is then $D$-dimensional
(in the physical case three-dimensional) below $\La_t$ and
described by \Eqn{II:Hitemp}.
We will use the results of Ginsparg's perturbative calculation
\cite{Ginsparg} of the effect of eliminating the non-zero modes
to approximately calculate the resulting ``boundary condition''
potential at $\La_t$. Note that no infra-red difficulties
arise in integrating out the $n\not=0$ modes as their
propagators are cut off by large effective masses $m^2_n=(2\pi nT)^2$.
Also note that we should not take $\La_t$ too much smaller than $T$
or else the couplings of the effective Lagrangian will
flow significantly due to fluctuations in $\phib_{i,0}(\kv)$,
for $k>\La_t$. Note that if greater accuracy is required, it is
(numerically) straightforward to perform the correct procedure outlined
above.

\hide{
A reasonable approximation \cite{TW} is to define the threshold scale
$\La_t = (K_{D-1}/K_D)T$ at which the $\La\gg T$ \eqn{II:Kaw}
and $\La\ll T$ \eqn{II:Hitemp} approximations agree.
We will then use the large $\La$ approximation for $\La > \La_t$,
and the small $\La$ approximation for $\La<\La_t$.
It is  then simplest to set $\La_0=\La_t$, and flow down using
\eqn{II:Kaw} to impose zero temperature renormalization conditions,
and flow down using \eqn{II:Hitemp} to obtain the finite temperature
coarse-grained effective potential.
}

\section{Application to the cubic anisotropy model}

After this long preamble we are finally ready to
calculate the appropriate initial three-dimensional
action at $\La_t$ for our toy system, and then give a general
account of its fluctuation-driven first-order
phase transition. Detailed analysis will be
given in the next section.

We are working at finite temperature
$T$, which enters the problem in two ways.
Firstly, as we mentioned above,
we rescale the fields by a factor of $\sqrt{T}$ so that all couplings in
the action have their usual engineering dimension
for a three-dimensional field theory, the four-dimensional
couplings being modified by factors of $T$.
The mass term is unaffected, but the $\phi^4$ couplings
are redefined
\beql{III:gdef}
g_i(\La)\equiv h_i(\La) T,
\eeql
where $h_i$ are the dimensionless (4D) couplings.

Second, the elimination of the $n\not = 0$ modes shifts the values
of the couplings $\mu^2$ and $h_i$ in \Eqn{I:model},
and also introduces higher dimension terms. However,
only the correction to the value of $\mu^2$ is not suppressed
by powers of the small couplings $h_i$ \cite{Ginsparg}.
For instance the shift in
the $\phi^4$ terms goes like $h_i^2$ -- suppressed by one power of $h_i$.
For our model the correction to the mass term is approximately
\beql{III:mshift}
m^2(\La_t) = - \mu^2 + { (N+2)h_1 + 3h_2 \over 72}T^2
\eeql
where as we will see the first-order transition temperature
$T_c$ is very close to the temperature where $m^2(T)=0$.

Our project is therefore to analyse the phase transitions of
the $N$ component scalar field theory defined by its
truncated potential $V^{\rm tr}$ at $\La_t$
(Eqs.~\eqn{III:trunc}, \eqn{III:mshift})
whose flow into the IR is given by \Eqn{II:Hitemp}.
We will now follow stages 1 and 2
of the procedure suggested at the end of Sec.~2.1,
first examining the flow of $V^{\rm tr}$, then evaluating the
full scale-dependent potential.

The ``truncated potential'', whose flows we study, is
\beql{III:trunc}
V^{\rm tr}_l(\phib) = \half m^2(l) \phib\cdot\phib
          +  {g_1(l)\over 4!} (\phib\cdot\phib)^2
               + {g_2(l)\over 4!} \sum_{i=1}^{N} (\phib_i)^4,
\eeql
where we have replaced the coarse-graining scale $\La$ with
a dimensionless variable $l$,
\beql{III:ldef}
\La=\exp(-l) \La(0),
\eeql
so $l$ goes to infinity in the IR, and we choose our units of energy
such that $\La(0)\equiv \La_t = 1$.
\hbox{} From the high temperature KIG equation \eqn{II:Hitemp}
we obtain the flow equations
\beql{III:gflow}
\ba{rcl}
\dsp {dm^2\over dl} &=& \dsp {K_D \e^{-Dl} \over \e^{-2l} + m^2}\,
{1\over 6} \Bigl( (N+2) g_1 + 3g_2\Bigr)~,\\[2ex]
\dsp {dg_1\over dl} &=& \dsp -{K_D \e^{-Dl} \over (\e^{-2l} + m^2)^2}\,
	\Bigl( g_1 g_2 + (N+8)g_1^2/6\Bigr)~,\\[2ex]
\dsp {dg_2\over dl} &=& \dsp -{K_D \e^{-Dl} \over (\e^{-2l} + m^2)^2}\,
	\Bigl( 2 g_1 g_2 + (3/2)g_2^2\Bigr)~.
\ea
\eeql
A more useful set of variables is obtained by rescaling by the
floating cutoff,
\beql{III:dimless}
\ba{rcl}
r(l)&\equiv & m^2(l)/\La(l)^2,\\[1ex]
\la_i(l)&\equiv& g_i(l)/\La(l)^{(4-D)}.
\ea
\eeql
These are the variables in which a second order phase
transition is described by a fixed point. Writing $(4-D)\equiv\ep$,
the RG equations for $\{ r,\la_1,\la_2 \}$ are:
\beql{III:lamflow}
\ba{rcl}
\dsp {dr\over dl} &=& \dsp 2r(l) + {K_D\over 1 + r}\,
{1\over 6} \Bigl((N+2)\la_1 + 3\la_2\Bigr)\\[2ex]
\dsp {d\la_1\over dl} &=& \dsp \ep\la_1(l) - { K_D\over (1 + r)^2}\,
	\Bigl(\la_1 \la_2 + (N+8)\la_1^2/6\Bigr)\\[2ex]
\dsp {d\la_2\over dl} &=& \dsp \ep\la_2(l) - { K_D\over (1 + r)^2}\,
	\Bigl(2 \la_1 \la_2 + (3/2)\la_2^2\Bigr).
\ea
\eeql
Note that this is where we introduce the $(4-\ep)$-expansion.
The physical value is $\ep=1$, but we will soon find it
necessary to take $\ep$ small in order to obtain any sensible results.
For $r\ll 1$ we find four fixed points of \eqn{III:lamflow}
(taking $N>4$ for the sake of simplicity -- the $N<4$ case
is entirely similar); they are (see Fig.~1),
\begin{enumerate}
\item The doubly unstable Gaussian fixed point at $\la_1=0$, $\la_2=0$.
\item The doubly stable Wilson-Fisher fixed point at
      $\la_1= 2\ep/(NK_D)$, $\la_2= 2(N-4)\ep/(3NK_D)$.
\item The Ising fixed point at $\la_1=0$, $\la_2=2\ep/(3K_D)$.
\item The Heisenberg fixed point at $\la_1=6\ep/((N+8)K_D)$, $\la_2=0$.
\end{enumerate}
The last two are both unstable in one direction in
the $\{ \la_1,\la_2 \}$ subspace.

As we will see below, first-order transitions occur when the
renormalization group flows take the couplings out of the region
where the quartic part of the potential is positive definite.
There are two types of instability:
\begin{enumerate}
\item Hypercubic diagonal instability. If $\la_1<0$ then
the potential is minimized for $\phib$ in a diagonal
direction, namely $\phib = (\pm 1,\pm 1,...,\pm 1)\phi/\sqrt{N}$
(of magnitude $\phi$). The quartic part of the potential is then
negative for $\la_1 + \la_2/N <0$.
\item Hypercubic axis instability. If $\la_2<0$ then
the potential is minimized for $\phib$ along
an axis, namely $\phib=(0,...,1,...,0)\phi$. The quartic
part of the potential is now negative for $\la_1 +\la_2 <0$.
\end{enumerate}
The unstable regions are indicated by shading in Fig.~1.

Solving the flow equations \eqn{III:lamflow} with $r\ll 1$ yields
the flow pattern shown in Fig.~1. Note that there are regions
of the coupling constant space that define sensible models (\ie that
are bounded from below), that have flows that {\it inevitably take
the couplings across the stability lines}, and into regions
without fixed points. For $N>4$ these regions are the two wedges
inside the stability region, but outside the quadrant $\la_1\geq 0$,
$\la_2\geq 0$ which is the domain of attraction of the Wilson-Fisher
fixed point. Systems with $\{\la_1(0),\la_2(0)\}$
in these wedges will undergo fluctuation-induced first-order phase
transitions.
Since the analysis of the phase transition for the two wedges
is nearly identical, we will choose to concentrate on the lower wedge
where $\la_2<0$, for which the RG flows take
$\la_i(l)$ across the stability line $\la_2(l)=-\la_1(l)$.
In this case the phase transition is from a disordered
phase to one with ordering along a hypercubic axis in $\phib$-space.

Also note that there exist
regions with arbitrarily small initial couplings ($\la_i\ll \ep$
at $\La_t$), that are first drawn towards the
unstable fixed points, where the couplings are of order $\ep$, before
veering away and then crossing one of the stability lines.
{\it Weakly} first-order transitions correspond precisely
to such values of the initial couplings where the amount of flow
is large.
Of course, there are also initial couplings that flow very little
before crossing these lines, never becoming very large. These values
correspond to strongly first-order transitions -- for which it is
not necessary to employ the RG to get a reasonable description of
the transition, at least on infinite length scales.
Ideally we would like to investigate the solution to the flow equations
in the case $\ep=1$. However the closed set of flow equations
\eqn{III:lamflow} was derived by truncation of
an otherwise infinite set of coupled equations for the
couplings in an arbitrary potential. Such a truncation is under
control if the theory is weakly coupled for {\it all scales}
$\La < \La_t$. For the weakly first-order case at $D=3$
the couplings would quickly flow
into a region of strong coupling where the loop-expansion
and the closely associated truncation fails.
This is the reason why a $4-\ep$ expansion is necessary if we
are to describe in a controlled way the weakly first-order case.

In any event, for any given $\{r(0),\la_1(0),\la_2(0)\}$, all three
of order $\ep$ or less,
we can follow the RG flows using \Eqn{III:lamflow}. Fixing $\la_1(0)$
and $\la_2(0)$, we can then fine-tune $r(0)$ so that the
completely coarse-grained potential $V_\infty(\phi)$ has two
degenerate minima. This means we have chosen the parameters such
that $T$ is the temperature of a first-order
phase transition. It turns out that we must tune $r(0)$
to be very close to a critical surface $r_c(\la_1,\la_2)$,
where $r_c\sim \O(\ep)$ (see \Eqn{IV:rc}).
As we evolve up in $l$, $r(l)$ stays close to the critical surface,
and hence the evolution follows the $r\ll 1$ trajectories of
Fig.~1. At a value of $l$ that we call $l_*$, we cross the stability line:
\beql{III:lstar}
\la_1(l_*)+\la_2(l_*)=0
\eeql
We will see in Section~4 that
in order to get the degenerate minima in $V_\infty$, we must have chosen
$r(0)$ such that
$r$ begins to deviate from $r_c$ at $l_*$.
($r(l_*)-r_c(\la_1(l_*),\la_2(l_*))\sim \O(\ep)$).
This deviation grows very fast,  and at $l\equiv l_f$
(see \Eqn{IV:lstar}) $r$ becomes of order $1$:
\beql{III:lf}
r(l_f) = 1 + \O(\ep),\qquad l_f= l_* + \half\ln(1/\ep) + \O(1)
\eeql
At this point the anomalous scaling of $r$ (the second term
in the first equation of \eqn{III:lamflow}) is suppressed
relative to the canonical scaling (the first term) by a factor
of $\ep$. Since the anomalous term in
the $r$ flow equation  has an factor of $1+r(l)$ in the
denominator, it is now doomed: $r$ grows like $\exp(2l)$
for all $l\gapp l_f$.
The $\la$'s suffer the same fate -- their anomalous scaling
is suppressed by powers of $1+r(l)$ too,
so for $l>l_f$, $\{r,\la_1,\la_2\}$ flow canonically
(or equivalently the physical parameters $\{m^2,g_1,g_2\}$
are independent of $l$). This is illustrated in Fig.~1 by
the straight line flow of the couplings
$\{\la_1,\la_2\}$ in the ``unstable'' regions.
It does not matter that the $\la$'s will
ultimately become large: the much faster growth of $r$
suppresses one and higher-loop effects completely,
and we can calculate reliably.
These flows can then be substituted into the RHS of \eqn{II:Hitemp}
to determine $V_l$ for any $l$. This is the topic to which we now turn.

If we take
$\phib$ to have magnitude $\phi$ along a hypercubic axis, the
quadratic and quartic parts of the renormalized potential are
\beq
V^{\rm tr}_l(\phi)= \half m^2(l)\phi^2
 + {\txt{1\over 24}}\Bigl(g_1(l)+ g_2(l)\Bigr) \phi^4~,
\eeq
expressed in dimensionful variables.
When we substitute this into the RHS of \eqn{II:Hitemp}
we get an integral expression for the higher-dimension terms
that are generated as a result of the coarse-graining process:
\beql{III:Effpot}
V_l(\phi) = V^{\rm tr}_l(\phi) + \De V_l\Bigl[ (g_1(s) +
 g_2(s)) \phi^2\Bigr] + (N-1)\De V_l\Bigl[ \third g_1(s) \phi^2\Bigr],
\eeql
where, again we have specialized to an axis direction.
Here $\De V$ is a functional, whose argument is a function of
$s$ over the range $0$ to $l$:
\beql{III:DeltaV}
\ba{rcl}
\De V_l[w(s)] &\equiv & \dsp \half K_D
\int_0^l\! \e^{-(D-\ep)s} ds\biggl\{ \ln\biggl[ 1 + {w(s)\over P(s)} \biggr]
- {w(s)\over P(s)} + \half \biggl({w(s)\over P(s)}\biggr)^2 \biggr\}\\[2ex]
P(s) &\equiv & \e^{-2s} + m^2(s).
\ea
\eeql
Note that we have subtracted off the quadratic and quartic parts of
$\De V_l$, since these constitute the truncated potential, whose
running we have already calculated.

Iterating the KIG equation once again (before taking $\phib$
to lie along the axis) we would find that the couplings of
the induced higher-dimension terms change the RG equations
for two and four-point terms. This change is perturbatively
small when the four-point coupling is itself small. As we
noted above, this is
in general only the case when we work
in a $(4-\ep)$-expansion with $\ep$ small.

\section{The phase transition}

We now give the detailed proof of the qualitative results
sketched above, and calculate the infinitely coarse-grained ($l=\infty$)
effective potential near the phase transition, finding what
value of $m^2(0)$ yields degenerate minima for given $g_1(0), g_2(0)$.

\subsection{The solution of the RG equations}

In view of the discussion at the end of Section~3.1, we only need
solve the flow equations for $l<l_f$, for which the $\la$'s
are always of order $\ep$.
Ultimately we will show  that the phase
transition occurs at values of $m^2(0)$ {\it etc.} such that
\eqn{III:lf} is satisfied.

We start with the flow of $r$.
Defining
\beql{IV:rc}
r_c(l)\equiv -{K_D \over 12} \Bigl( (N+2)\la_1(l) + 3 \la_2(l) \Bigr),
\eeql
(note that $r_c\sim\O(\ep)$) we can rewrite the flow of $r$ as
\beq
{d r \over d l} = 2r - {2r_c\over 1+r} + \O(\ep^2).
\eeq
As long as $r\sim\O(\ep)$ we can ignore the $(1+r)$ denominator;
When $r\sim\O(1)$ the whole second term is subleading. Thus
\beq
{d r \over d l} = 2r - 2r_c + \O(\ep r)\quad \forall l\lapp l_f
\eeq
Now $dr_c/dl\sim \O(\ep^2)$, so to leading order in $\ep$,
\beql{IV:rsol}
\ba{rcl}
r(l) &=& r_c(l) + \bar t \e^{2l}\\[1ex]
\hbox{\ie}\quad m^2(l) &=& \e^{-2l}r_c(l) + \bar t\\[1ex]
\bar t &=& \e^{-2l_f}(1 + \O(\ep))
\ea
\eeql
Putting this into words, we have found that there is an unstable
``critical surface'' $r_c(l)$. If $r$ starts off on the critical
surface ($\bar t=0$) then it never leaves,
its flow within the surface being
determined by the flow of  the $\la$'s.
In order to see a subsidiary minimum
in the potential we will have to choose $\bar t$ exceedingly small,
so that $l_f$ is large. Specifically, we must ensure that
$r$ does not grow to $\O(1)$ until (just) after the couplings
cross the stability line, \ie at $l_* + \O(\ln(1/\ep)$.

Our next task is to solve the flow of $g_1$ and $g_2$,
or equivalently $\la_1$ and $\la_2$. For $l<l_*$, \eqn{III:gflow}
becomes
\beq
\ba{rcl}
\dsp \ep{dg_1\over dx} &=& \dsp -{K_D \over 6}
	\Bigl( 6 g_1 g_2 + (N+8)g_1^2\Bigr)~,\\[2ex]
\dsp \ep{dg_2\over dx} &=& \dsp -K_D
	\Bigl( 2 g_1 g_2 + (3/2)g_2^2\Bigr)~,\\[2ex]
\hbox{where}\quad x &=& \exp(\ep l)
\ea
\eeq
It is convenient, following Rudnick \cite{Rud}, to define
\beql{IV:fdef}
F\equiv -{g_1\over g_2},
\eeql
($F$ is positive for the case we consider) so that
\beq
{dg_1 \over dg_2} =  {(N+8)/3 F^2 - 2F \over 3-4F},
\eeq
and
\beq
g_2 {dF \over dg_2} = {dg_1\over dg_2} + F = {(N-4)F^2/3 +F \over 3-4F }
\eeq
which is readily integrated to yield
\beql{IV:g2sol}
g_2(l) = g_2(0) \biggl({F(0)\over F(l)}\biggr)^3
\biggl({ (N-4)F(l) +3 \over (N-4)F(0)+3}\biggr)^{3N/(N-4)}
\eeql
Now all we have to do is find $F$. By definition \eqn{IV:fdef} and the
RG equations \eqn{III:gflow},
\beq
{dF\over dx} = {K_D g_2 \over 6\ep} \Bigl( 3F + (N-4)F^2\Bigr)
\eeq
so by \eqn{IV:g2sol},
\beql{IV:feqn}
\ba{rcl}
\dsp {d F \over dx} &=& \dsp {A\over F^2} ( 3+(N-4)F)^{\al+1}\\[2ex]
A &\equiv & \dsp {K_D g_2(0) \over 6\ep} {F(0)^3
   \over (3+(N-4)F(0))^\al}\\[2ex]
\al &\equiv & 3N/(N-4)
\ea
\eeql
Note that $A$ is negative.
Solving, we find the dependence of $F$ on $l$:
\beql{IV:fsol}
\ba{rcl}
A (N-4)^3\,(\e^{\ep l} -1) &=& \dsp \Bigl\{Z\Bigl(3+(N-4)F(l)\Bigr)
   - Z\Bigl(3+(N-4)F(0)\Bigr)\Bigr\}\\[1ex]
Z(y) &\equiv & \dsp {y^{2-\al}\over 2-\al} - {6y^{1-\al}\over 1-\al}
 - {9y^{-\al}\over\al}
\ea
\eeql
Armed with Eqs.~\eqn{IV:g2sol} and \eqn{IV:fsol} we can now
obtain $g_1(l)$ and $g_2(l)$ for any $l<l_*$
given $g_1(0)$ and $g_2(0)$.

\subsection{The effective potential near the phase transition}
We have managed to obtain the flows in closed form. Stage 2 of
our procedure now calls for us to substitute them in to
\Eqn{III:Effpot} and send $l\to\infty$ to get the effective
potential, the form of which will tell us
whether there is a first-order phase transition.
In order to obtain understandable (and self-consistent)
results we must again make the approximation of keeping
only the leading order non-trivial terms in the $\ep$-expansion.

We have previously argued (below \Eqn{III:lf}) that $r$, $\la_1$,
$\la_2$ scale canonically for $l>l_f$,
which means that $m^2$, $g_1$, and $g_2$ can be taken to be
constant. Thus the old non-RG-improved effective potential
formula is valid in this region. It is therefore convenient
to define the following variants of $\De V$ (see \Eqn{III:DeltaV})
where the couplings are evaluated at some particular $l$ and do not flow:
\beql{IV:IR-UV}
\ba{rcl}
\De U^{IR}_l(w) &\equiv& \dsp \half K_D
\int_l^\infty\! \e^{-(D-\ep)s} ds
  \ln\biggl[ 1 + {w\over P(s)}\biggr]\\[2ex]
\De U^{UV}_l(w) &\equiv & \dsp \half K_D
\int_0^l\! \e^{-(D-\ep)s} ds\bigg\{ \ln\biggl[ 1 +{w\over P(s)} \biggr]
- {w\over P(s)} + \half \biggl({w\over P(s)}\biggr)^2 \bigg\}\\[2ex]
P(s) &\equiv & \e^{-2s} + m^2(l)
\ea
\eeql
so that the potential becomes
\beql{IV:Bigl}
\ba{rcl}
V_\infty =  V^{\rm tr}_{l_f}(\phi)
&+& \De V_{l_f}\Bigl[ \half(g_1(l) + g_2(l)) \phi^2\Bigr]
+ (N-1)\De V_{l_f}\Bigl[ \sixth g_1(l) \phi^2\Bigr]\\[1ex]
&+& \De U^{IR}_{l_f}\Bigl( \half(g_{1\f} + g_{2\f}) \phi^2\Bigr)
+ (N-1)\De U^{IR}_{l_f}\Bigl( \sixth g_{1\f} \phi^2\Bigr)\\[1ex]
g_{1\f} \equiv g_1(l_f),\quad\hbox{etc.}
\ea
\eeql
The first three terms represent the ``ultraviolet'' contribution,
which requires integrating over the flowing couplings from
$l=0$ to $l_f$.
The last two terms are the ``infrared'' contribution, which
are easy to evaluate, since nothing flows in the integrand.

Actually, to the leading non-trivial order in $\ep$ to which we
work, one can also avoid
the complications of integrating over flowing mass and couplings
in the expression for $\De V$ (\Eqn{III:DeltaV}),
which is integrated from $l=0$ to $l_f$.
To see this, first note that the flow of the mass in $\De V$
can be ignored, since it only enters through the  propagator
$P(l)=\exp(-2l)+m^2(l) = \exp(-2l)(1+\O(\ep)) + \bar t$,
so we can set $m^2(l)=m^2(l_f)=\bar t$ to leading order in
$\ep$, with $\bar t$ given in \Eqn{IV:rsol}. Note that
$\De V$ only supplies the $\phi^6$ and higher terms -- the
flow of $m^2$, $g_1$, $g_2$ is explicitly taken into account
in the lower order terms in $V^{\rm tr}$.
To see that the flows of $g_1$ and $g_2$ can be ignored, look
a typical term in $\De V_{l_f}$:
\beq
\half K_D \int_0^{l_f} {\e^{-(D-\ep)s} w(s)^n \over (\e^{-2s}
+ \bar t)^n}\,ds
\qquad (n\geq 3)
\eeq
where $w(s)\sim g(s)\phi^2$.
Because $\bar t=\exp(-2l_f)$ this integral is dominated by the region
close to $l_f$: the domain $l_f-\half\ln(1/\ep)<l<l_f$ gives
the dominant contribution to order $\ep$.
In this relatively small range, the $g$'s hardly flow
at all, since $(1/g)dg/dl \sim \ep$,
so we can take them to have their
values at $l_f$ throughout the whole range. Thus
$\De V_{l_f}[w(s)] = \De U_{l_f}^{UV}(w(l_f))$ to lowest order in
$\ep$, and
\beql{IV:Effpot}
\ba{rcl}
V_\infty(\phi) =  V^{\rm tr}_{l_f}(\phi)
&+& \De U^{UV}_{l_f}\Bigl( \half(g_{1\f} + g_{2\f}) \phi^2\Bigr)
+ (N-1)\De U^{UV}_{l_f}\Bigl( \sixth g_{1\f} \phi^2\Bigr)\\[1ex]
&+& \De U^{IR}_{l_f}\Bigl( \half(g_{1\f} + g_{2\f}) \phi^2\Bigr)
+ (N-1)\De U^{IR}_{l_f}\Bigl( \sixth g_{1\f} \phi^2\Bigr)
\ea
\eeql
We have now expressed the effective potential entirely in terms of
simple integrals that depend on the relevant couplings at $l_f$.

Performing the integrals, we find that
\beql{IV:DeltaU}
\ba{rcl}
\De U_l(\wb) &\equiv &\De U^{IR}_l(w)+\De U^{UV}_l(w) \\[1ex]
& =& \dsp {K_D \over 8}\e^{(\ep-4)l}
\biggl\{  2\wb  - {r_l\over 1+r_l} \wb^2
 + f(r_l +  \wb) - f(r_l)
- (\wb^2 + 2r_l\wb) \ln(1 + r_l)\biggr\}\\[2ex]
f(x) &\equiv & x^2\Bigl(\ln(x)-\half\Bigr),\qquad r_l \equiv
r(l),\qquad \wb \equiv \e^{2l} w,
\ea
\eeql
and defining a rescaled field
\beql{IV:scaledphi}
\phb^2_f \equiv \phi^2/\La(l_f)^{(2-\ep)}=\e^{(2-\ep)l_f} \phi^2,
\eeql
we find that the potential is
\beql{IV:Potlf}
\ba{rcl@{}l}
V_\infty(\phb) &=& \dsp \exp[(\ep-4)l_f] \biggl\{ &
\half r_{l_f}\phb^2 + \twfourth (\la_{1\f}+\la_{2\f})\phb_f^4
 + \De U_{l_f}\Bigl( \half(\la_{1\f} + \la_{2\f})\phb_f^2 \Bigr)\\[1ex]
&&&+ (N-1)\De U_{l_f}\Bigl( \sixth\la_{1\f}\phb_f^2 \Bigr)\biggr\}\\[1ex]
\la_{1\f}&\equiv & \la_1(l_f),\quad\hbox{etc.},&\qquad r_{l_f}=1.
\ea
\eeql

We are now almost finished. We can now see for what values of
$\la_{1\f},\la_{2\f}$ there is a phase transition at this temperature.
We look for a subsidiary minimum
at non-zero $\phb_f$, degenerate with the one at the origin.
Anticipating the final result, we guess that it will occur for
\beql{IV:Orders}
\ba{rcl}
\la_{1\f}, \la_{2\f} &\sim & \O(\ep),\\[1ex]
(\la_{1\f}+ \la_{2\f}) &\sim & \O(\ep^2 \ln \ep)\\[1ex]
\phb_f &\sim &\O(1/\ep)
\ea
\eeql
Using this in \eqn{IV:Potlf} and keeping only terms of order
$(1/\ep^2)\ln\ep$ or $1/\ep^2$, we find
\beql{IV:Vf}
V_\infty(\phb_f)= \half\phb_f^2 + \twfourth(\la_{1\f}+\la_{2\f})\phb_f^4
+ {K_D (N-1)\over 8} (\sixth\la_{1\f}\phb_f^2)^2
\Bigl(\ln(\twelfth\la_{1\f}\phb_f^2) - 1\Bigr)
\eeql
which has the required degenerate subsidiary minimum for
\beql{IV:Lamsum-lf}
\la_{1\f}^\pt + \la_{2\f}^\pt = -{K_D (N-1)\over 12} (\la_{1\f}^\pt)^2
\ln\biggl[{12\over \la_{1\f}^\pt K_D (N-1)}\biggr],
\eeql
(which is of order $\ep^2\ln\ep$, as expected), at
\beql{IV:Phi-lf}
(\phb_f^\pt)^2 = {144 \over \la_{1\f}^2 K_D (N-1) },
\eeql
(which is of order $1/\ep^2$ as expected).
This demonstrates that indeed our model possesses a first-order
phase transition of the type claimed.

We could at this stage perform the final task of
re-expressing these values, and the mass $r(l_f)=1$, in terms
of the initial renormalized parameters of the theory at
$\La_t$ (\ie $l=0$) using Eqs.~\eqn{IV:g2sol}, \eqn{IV:fsol},
and \eqn{IV:rsol}. Let us, however, delay this until our
comments and conclusions in Section~5.

It is also possible to express the infinitely coarse-grained
free-energy in terms of the couplings evaluated at $l_*$,
rather than $l_f$. The independence of the physical
results (to leading order in $\ep$) enables us to check that
we have not made an error in our calculation.
To do this we need to flow
$r_{l_f},\, \la_{1\f},\, \la_{2\f}$ up to $l_*$.
First we must find $l_*$, which is defined by
$\la_1(l_*) + \la_2(l_*) = 0$.
We therefore define
\beq
u(l)\equiv\e^{-\ep(l-l_*)}(\la_1 + \la_2)
\eeq
and we find, by Eqs.~\eqn{III:lamflow}, \eqn{IV:rsol},
to leading order in $\ep$,
\beq
{\p u\over \p l} = -\sixth K_D(N-1)\la_1^2(l)
  { \e^{-\ep(l-l_*)} \over (1 + \e^{2l}\bar t\,)^2 }
\eeq
We can integrate this equation from $l_f$ to $l_*$, since
$\la_1$ varies by an amount subleading in $\ep$
in this range. Under the correct assumption
that $\exp(l_f-l_*)\sim \O(1/\ep)$, we obtain
\beq
u(l_f) = \twelfth K_D(N-1)\la_{1*}^2 \Bigl(\half
   + \ln[2 \e^{-2(l_f-l_*)}]+\cdots \Bigr).
\eeq
Hence
\beql{IV:lstar}
\exp(2(l_f-l_*)) = {24\e^{1/2}\over \la_{1*} K_D(N-1)}
\eeql
which is much greater than 1 as anticipated.

We can now go back to Eqs.~\eqn{IV:Effpot} and \eqn{IV:DeltaU}
and re-express the potential as a function of quantities defined
at $l_*$, obtaining an expression like \Eqn{IV:Potlf} with
$f\to *$. Let us see for what mass and couplings we expect to
see the degenerate subsidiary minimum.
We flow the mass up to $l_*$ by using
\eqn{IV:rsol}, \eqn{IV:rc} and \eqn{IV:lstar}
\beql{IV:rstar}
r^\pt(l_*) = \twelfth K_D(N-1) \la_{1*} \Bigl(\half \e^{-1/2} -1\Bigr).
\eeql
The minimum will be at
\beql{IV:phistar}
(\phb_*^\pt)^2 = \e^{2(l_f-l_*)}(\phb_f^\pt)^2 = {6\e^{-1/2}\over \la_{1*}}.
\eeql
We conclude, analogously to \eqn{IV:Orders}, that the interesting
parameter range for the potential is
\beql{IV:Order*}
\ba{rcl}
r(l_*) &\sim & \O(\ep),\\[1ex]
\la_{1*}= \la_{2*} &\sim & \O(\ep),\\[1ex]
\phb_* &\sim &\O(\ep^{-1/2}).
\ea
\eeql
Keeping leading order in $\ep$ we find the potential
\beql{IV:Vstar}
V_\infty(\phb_*) = \half r(l_*)\phb_*^2 + {K_D(N-1) \over 8}
\Bigl\{ \third \la_1 \phb_*^2 + (\sixth\la_1 \phb_*^2)^2
\Bigl[\ln(\sixth\la_1 \phb_*^2) - \half\Bigr]\Bigr\}
\eeql
This should be exactly the same as \eqn{IV:Vf}, and in fact
it is. It has a degenerate subsidiary minimum at
the values of $r$ and $\phb$ predicted by \eqn{IV:rstar} and
\eqn{IV:phistar}.
Finally, note that our result is identical (to leading order in
$\ep$ and after suitable redefinitions) with that of Rudnick
\cite{Rud}, who previously analyzed a Landau-Ginzburg free energy
of the form \Eqn{I:model} within a condensed-matter context.

\section{Comments and conclusions}

We now wish to make some comments concerning the formalism
and calculation presented above.

1) We have shown that it is possible to systematically calculate
a {\it scale-dependent} coarse-grained free energy that describes
a (fluctuation-induced) first-order phase transition. As we argued
in the introduction, if one wishes to explicitly follow
the {\it dynamics} of the phase transition (such as critical bubble
nucleation and expansion) then one must coarse-grain not out to
infinity ($\La\to 0$), but to some appropriate length
scale. We want to re-emphasize that this procedure should
be familiar from the study of effective gauge theories -- there one
integrates out the heavy degrees of freedom to get the
effective Lagrangian describing the dynamics of the light modes.
In our case we might want, for instance, to accurately calculate
bubble wall properties -- we should
therefore stop coarse-graining when we reach the bubble-wall thickness,
and use the coarse-grained free energy at that length-scale.

As an illustration of this procedure we can look at the (typical)
case where this thickness is the same order as the correlation length.
In terms of the variable $r_l$ used in Section~4 this is precisely
when $r_l=1$, \ie $l=l_f$. We therefore need $V_{l_f}$. However this is
nothing more than the expression for the free-energy in \Eqn{IV:Bigl}
without the extra infrared contributions
\beql{V:Vatlf}
V_{\infty}(\phi) - V_{l_f}(\phi) =
\De U^{IR}_{l_f}\Bigl( \half(g_{1\f} + g_{2\f}) \phi^2\Bigr)
+ (N-1)\De U^{IR}_{l_f}\Bigl( \sixth g_{1\f} \phi^2\Bigr),
\eeql
where we recall from \eqn{IV:IR-UV} that
\beql{V:UIR}
\De U^{IR}_l(w) \equiv \dsp \half K_D
\int_l^\infty\! \e^{-(D-\ep)s} ds
  \ln\biggl[ 1 + {w\over P(s)}\biggr].
\eeql
In general, the difference between $V_{\infty}(\phi)$
and $V_{l_f}(\phi)$ (given for
our toy model by \Eqn{V:Vatlf}) can be quite significant.
In our model it turns out that within the $(4-\ep)$-expansion, and
for the orders of magnitude of $\phi$, and the couplings, of interest
\Eqn{IV:Orders}, this difference is actually subleading by one
power of $\ep$. Since we have not calculated our RG flows to higher
order in $\ep$ we therefore cannot give a reliable explicit
expression for the values of the integrals in \eqn{V:Vatlf}.
Of course, despite the fact that they are subleading in $\ep$,
at $\ep=1$ their contribution can be important. We expect that
for the gauged case we describe below the difference between
$V_{\infty}(\phi)$ and $V_{l_f}(\phi)$ will not be subleading
in $\ep$.

2) We now want to return to the one loose-end
of Section~4 -- namely the re-expression of
the properties of the potential in terms of the initial
variables at $l=0$. Using the solutions, Eqs.~\eqn{IV:g2sol},
\eqn{IV:fsol}, and \eqn{IV:rsol}, we derived in Section~4.1
to the RG equations this is a straightforward task.
However the expressions one gets in
doing this are quite lengthy.
A simplification occurs if we take the initial ratio
of the couplings $F_0 = -g_1(0)/g_2(0) \gg 1$. For example,
in this limit the value of the field at the new
(true vacuum) minimum is given by
\beql{V:phiatmin}
\phi^2 = \e^{-(2-\ep)l_*}{N(N+2)(N+8)\over N(N+5)+3}{ K_D \over \ep},
\eeql
where
\beql{V:e2l}
\e^{-(2-\ep)l_*} = \biggl\{ {N(N+2)(N+8)\over N(N+5)+3}
{ g_1(0) K_D \over 6\ep}
\biggl( { N-1 \over N-4 }\biggr)^{3N/(N-4)} \biggr\}^{(2-\ep)/\ep}
F_0^{-\txt {(2-\ep)(N+8)\over \ep (N-4)}}.
\eeql
($l_*$ is the value of the flow parameter at which
the system crosses the stability line $g_1(l_*) + g_2(l_*)=0$).
As advertized in the Introduction
these expressions are quite non-analytic as a function
of the initial couplings and $\ep$.
The expression for $\phi^2$ \Eqn{V:phiatmin} also exemplifies the
connection between weakly first-order transitions and
significant flow in $l$ that we mentioned in Section~3.
This is because when $l_*$ is large
the value of $\phi^2$ in the new minimum (at the transition
temperature) given by \eqn{V:phiatmin} is small compared to its
zero-temperature value.

3) In the introduction we motivated the study of our toy scalar
model by stating that it shared many features in common with
the phase transition occurring in gauged systems. One way of
understanding why this is true is to look at the RG
flow equations (once again in $(4-\ep)$-spatial dimensions)
for the couplings of the prototypical gauged
system -- the $N$-component Abelian Higgs model.
Again we have two couplings that are marginal near four-dimensions,
the quartic scalar self-coupling $h(l)$, and
the gauge coupling $e^2(l)$. Their RG equations
are
\beql{V:gaugeRG}
\ba{rcl}
\dsp {d e^2\over dl} &=& \dsp \ep e^2 - {N K_D \over 3} e^4 \\[2ex]
\dsp {dh\over dl} &=& \dsp (\ep-2\eta)h - {(N+4)K_D\over 2(1+r)^2} h^2
                   -12K_D e^4,
\ea
\eeql
where $\eta=-3K_D e^2/(1+r)$ is the anomalous dimension of the
scalar field induced by wavefunction renormalization, and $r(l)$
is the (dimensionless) scalar mass parameter with a RG equation
we shall not display. (As in Section~3 we have implicitly
included the appropriate factor of temperature $T$ induced by the
rescaling of the fields, as well as then dividing out by $\La(l)^\ep$
so as to get dimensionless couplings.) For any reasonable number
of complex scalar fields $N<366$ these equations have
only two fixed points -- the doubly unstable Gaussian fixed point
at the origin, and a fixed point at
\beql{V:gaugeFP}
h = {2\ep\over (N+4)K_D},\qquad  e^2 = 0,
\eeql
which is unstable in the direction of the gauge coupling. The
qualitative pattern of the flows in the $\{h, e^2\}$ plane
is illustrated in Fig.~2.

The important point is that there are well-defined models (indeed
the whole of the $e^2>0$, $h>0$ quadrant!) whose
flows inevitably take the system outside the naive stability
region (the same $e^2>0$, $h>0$ quadrant). Just as in the scalar
case we have considered above, this implies that the system undergoes
a fluctuation induced (finite-temperature) phase transition.
Furthermore, the work of Ginsparg \cite{Ginsparg} shows that the
standard model and almost all other gauge systems of interest
fall into this class. In a future publication we will show
how the properties of these first-order phase transitions can
be analyzed by the application of the Wilson RG and the
$(4-\ep)$-expansion.

4) In a recent very interesting work, Shaposhnikov \cite{Shaposh}
argued that to fully understand the finite-temperature
phase transition that occurs in gauge theories, it is necessary
to include {\it non-perturbative} contributions to the free-energy.
This, he argued, is because the effective three-dimensional
theory of the zero Matsubara modes is a confining theory
in the far infrared -- at least in the non-Abelian case.
It was also argued that the non-perturbative effects can
make the transition in, say, the minimal standard model
considerably stronger for a given Higgs mass than a perturbative
analysis would indicate.
It is interesting to see what our approach says about this
general question.

We have already noted that
the effective $(4-\ep)$-dimensional theories that we must
consider if we want to study the phase transition
have the property that
as we flow into the infra-red the effective
couplings first flow towards the unstable fixed point,
which for $\ep=1$ is well into the strong-coupling regime.
The same phenomenon
occurs in gauge theories, as exemplified by the RG flows of the
Abelian Higgs model illustrated in Fig.~2.
One is therefore tempted to say that this is almost a
proof of the correctness of Shaposhnikov's suggestion
at $\ep=1$. {\it All} the theories are
in their strong-coupling regime and non-perturbative
effects are important. However, this does not necessarily
mean that we cannot calculate.

To see this,
consider the flows in Fig.~1 that take us towards
the {\it stable} Wilson-Fisher fixed point at
$\la_1= 2\ep/(NK_D)$, $\la_2= 2(N-4)\ep/(3NK_D)$, corresponding
to a second-order transition.
For $\ep=1$ this is also far into the strong-coupling region,
so surely we must consider non-perturbative effects
if we are to get accurate predictions for the properties
of this transition (such as the critical exponents).
In fact, the experimentally
observed critical exponents are amazingly well
described by an expansion up to only $\O(\ep^2)$.
Even more accurate results can be obtained by
Lipatov techniques for estimating large
order behavior of the $\ep$-expansion together with Borel resummation.
These techniques are closely related to the usual
semi-classical methods employed in dealing with weakly non-perturbative
effects. Thus we might say that for this second-order case it seems
to be that the non-perturbative effects that are present are
fairly benign -- and can be handled within more sophisticated
versions of the $\ep$-expansion.

What about the first-order case -- especially for gauge
theories? Here the situation is less clear, mainly as
a consequence of the relatively limited amount of work
that been performed so far. There has been for example, to
our knowledge, no extension of the Lipatov techniques to the
first-order case. One obvious difference
between the first and second-order cases is that in terms
of the dimensionless variables, $\la_i(l)$, the flows
take one out to ``infinity'' in
coupling constant space, rather than to a fixed point.
This is actually benign in the scalar
case since soon after we cross the stability line we reach
a point, $l_f$, where the increase in the scalar mass
cuts off any further non-canonical scaling of the couplings,
so the $\la_i(l)$ move along a straight line trajectory
projecting out from the origin (see Fig.~1)
-- in other words the dimensionful couplings are to
a good approximation {\it constant} after $l_f$.
However, in the gauge case the flow of the gauge coupling
doesn't seem to be cut off in quite such a simple way. This is
consequence of the lack of any factors of $1/(1+r(l))^n$
appearing in the RG equation for $e^2$ in \eqn{V:gaugeRG}.
Non-Abelian theories also present new difficulties.
We are at present considering these issues.
In any case it is certainly true that the RG techniques we have
utilized in this paper lead to considerably greater insight into
this question than more standard diagrammatic resummation
techniques.

\vspace{36pt}
\centerline{\bf Acknowledgments}
We wish to thank Jacques Distler, Mark Goulian,
and especially, Raman Sundrum for helpful discussions.
JMR wishes to thank the Department of
Energy for the award of a DOE Distinguished Research Fellowship.
This work was also supported in part by
the U.S. Department of Energy under Contract
DE-AC03-76SF00098, and by the National Science Foundation.
MGA thanks LBL for hospitality and support as a summer visitor.

\appendix
\section{Derivation of the evolution equation}

In this appendix we will
derive the KIG evolution equation \eqn{II:Kaw} from Polchinski's
evolution equation \eqn{II:poleqn}, for the case of a single
scalar field $\phi$ in 4 dimensions. The generalization to
multiple fields and arbitrary dimension is straightforward.
We first show that, for momenta below the cutoff $\La$, Polchinski's
interaction Lagrangian $L_\La[\phi]$ is the same as the 1PI generating
functional $\Ga_\La[\phi]$ defined by Bonini, D'Attanasio, and
Marchesini (BDM) \cite{BDM} in their perturbative analysis of
Polchinski's equation. We then use their evolution equation for
$\Ga_\La[\phi]$ to derive the KIG evolution equation for the
effective potential.

To clarify matters we will take the momentum cutoff at $\La$ to be sharp,
and use superscripts $>$ and $<$ to distinguish degrees of freedom
above and below the cutoff respectively. (For a rigorous analysis see
Morris's very clear exposition
\cite{Mor}, in which the sharp cutoff limit is taken at the end.)
The propagator has two domains,
\beql{A:props}
\ba{rrcl}
\hbox{Infrared (IR):}& D_{<}(k) &=& K(k^2/\La^2)/(k^2 + m^2),\\[1ex]
\hbox{Ultraviolet (UV):}& D_{>}(k) &=& [1-K(k^2/\La^2)]/(k^2 + m^2).
\ea
\eeql
These add up to the full propagator $D(k)$. For most purposes
it is natural to take $K(x)$ to be a step function that is 1
for $x<1$ and zero otherwise. To make sense of BDM's definitions,
however, we will have to allow the UV propagator to have some
infinitesimal value in the IR,
and take the limit as it goes to zero.

The full generating functional of the theory can be split into
IR and UV portions:
\beql{A:genfunc}
\ba{r@{\,}c@{\,}l}
Z[J] &=& Z[J^<,J^>] =
\dsp \int [D\phi]
       \exp\biggl\{ \int_k \Bigl[-\half\phi(k)
       D^{-1}(k)\phi(-k) + J(k)\phi(-k)\Bigr] + L_{\rm bare}[\phi]
       \biggr\}\\[2ex]
&=&\dsp \int [D\phi^<]
       \exp\biggl\{ \int_k \Bigl[-\half\phi^<(k)
       D^{-1}_{<}(k)\phi^<(-k) + J^<(k)\phi^<(-k)\Bigr] + L_\La[\phi^<,J^>]
       \biggr\}.
\ea
\eeql
Comparing with \Eqn{II:gfunct} we see that Polchinski's interaction
Lagrangian $L_\La$ is the functional integral over
UV fields, with a given IR background field $\phi^<$ and
in the presence of a UV source $J^>$,
\beql{A:L}
\ba{r@{}l}
\exp L_\La[\phi^<,J^>] = \dsp\int [D\phi^>]
       \exp\biggl\{&\dsp \int_k \Bigl[-\half\phi^>(k)
       D^{-1}_{>}(k)\phi^>(-k) + J^>(k)\phi^>(-k)\Bigr]\\[2ex]
       & + \, L_{\rm bare}[\phi^<,\phi^>] \biggr\}.
\ea
\eeql
\Eqn{A:genfunc} tells us that we should think of $L_\La[\phi^<,0]$ as
the effective action ``felt'' by the IR fields.
We now want to show that it is the same as BDM's effective
action (for momenta below $\La$)
\beq
\Ga_\La[\psi^<,J^>\!=\!0] = \int_{k<\La} j^<(k)\psi^<(k) - W[j^<,0],
\qquad \hbox{with}\quad \psi^< = {\de W \over \de j^<},
\eeq
which is the Legendre transform of their connected generating
functional with UV propagator,
\beql{A:W}
\ba{r@{}l}
\exp W_\La[j^<,J^>] \equiv \dsp\int [D\phi]
    \exp\biggl\{ &\dsp \int_k \Bigl[-\half\phi(k)
    D^{-1}_{>}(k)\phi(-k) + j^<(k)\phi^<(-k) + J^>(k)\phi^>(-k)\Bigr]\\[2ex]
    & + \,L_{\rm bare}[\phi^<,\phi^>]  \biggr\}.
\ea
\eeql
A peculiar (but essential) aspect of this definition is that the
functional integral is over IR as well as UV fields, even though
the propagator is cut off in the IR. Moreover, there is an IR source
coupled to the IR field. This only makes sense if we give
$D_{>}$ some infinitesimal value $\epsilon/(k^2+m^2)$ in the IR, and
perform the IR part of the functional integral before
sending $\epsilon\to 0$. $D^{-1}_{>}(k<\La)$ is then very large, and
in combination with the $j^<\phi^<$ term  gives a $\de$-function,
requiring $\phi^<(k) =\psi^<(k) = D_{>}(k)j^<(k)$.
The result is
\beql{A:Ga}
\ba{r@{}l}
\exp -\Ga_\La & [\psi^<,0] =\dsp \exp\biggl( -\half \int_k \psi^<(k)
   D^{-1}_{>}(k)\psi^<(-k)\biggr)\\[2ex]
&\dsp\times \int [D\phi^>] \exp\biggl\{\dsp \int_k \Bigl[-\half\phi^>(k)
    D^{-1}_{>}(k)\phi^>(-k) + \Bigr]
    - L_{\rm bare}[\psi^<,\phi^>]  \biggr\}.
\ea
\eeql
We see that $\Ga_\La[\psi^<]$ is indeed the same as Polchinski's effective
action $L_\La[\phi^<]$ \Eqn{A:L}, except that it has an
additional inverse propagator
term in front. As we send $\epsilon\to 0$, eliminating BDM's
infinitesimal IR component of $D_{>}$, this term goes to
infinity ($D_{>}(k<\La)\sim\epsilon$). This is not a disaster---it
just reflects the fact that we in effect gave the IR modes infinite
mass in order to treat them as a background. It only affects the
2-point function, which BDM write as
\beq
\Ga^{(2)}_\La(k) = D_{>}^{-1}(k) + \Si_\La(k),
\eeq
and they obtain an evolution equation for the $\epsilon$-independent
part $\Si_\La$.
We therefore write
\beql{A:V}
\ba{rcl}
V_\La(\phi)&=&(m^2 + \Si_\La)\phi^2 + Y_\La(\phi),\\[1ex]
Y_\La(\phi)&=&\dsp\sum_{n=2}^\infty {1\over (2n)!} \Ga^{2n}_\La \phi^{2n},
\ea
\eeql
where the $\Ga^{2n}_\La$ are zero-momentum $n$-point 1PI functions.

We now begin the second stage of our calculation, showing that
BDM's evolution equation for $V_\La$ reproduces the KIG equation
\Eqn{II:Kaw}, in the approximation where momentum-dependence
in the action is ignored. Expanding $V_\La$ in the same way
as $\Ga_\La$, so that $V_\La^{(2)}=m^2+\Si_\La$
but for $n>1$, $V_\La^{(2n)}=\Ga_\La^{(2n)}$, BDM find
\beq
\La {\p V^{(2n)}_\La \over \p\La} = \half \int {d^4 q \over (2\pi)^4}
\La {\p D_\La \over \p\La}(q) {\bar\Ga^{(2n+2)}_\La(q) \over
  (1 + D_\La(q) \Si_\La(q))^2 },
\eeq
where we have written $D_\La(q)\equiv D_{>}(q)$, and
\beq
\bar\Ga^{(2n+2)}_\La(q) = \Ga^{(2n+2)}_\La - \sum_{l=1}^{n-1}
{2n\choose 2l} {\Ga^{(2l+2)}_\La \bar\Ga^{(2n-2l+2)}_\La(q) \over
D^{-1}_\La(q) + \Si_\La(q) },
\eeq
so that
\beql{A:messy}
\ba{r@{}l}
\dsp\La {\p V^{(2n)}_\La \over \p\La} = \half K_4 \sum_{N=1}^n
\Biggl\{I_N(\La)\sum_{l_1=1}^{n-1} \sum_{l_2=1}^{n-l_1-1}\cdots
   \sum_{l_{N-1}=1}^{n-l_1-\cdots-l_{N-2}-1} &\dsp
\biggl[ -
{ (-1)^N (2n)! \over (2l_1)!\cdots(2l_N)!} \\[3ex]
  &\times \Ga^{2l_1+2}_\La \cdots \Ga^{2l_N+2}_\La \biggr]
\Biggr\}.
\ea
\eeql
The momentum integral is contained in $I_N$,
\beq
I_N(\La) = \int \!q^3 dq\, \La {\p D_\La \over \p\La}(q)
{D^{-2}_\La(q) \over (D^{-1}_\La(q) + \Si_\La(q))^{N+1}}.
\eeq
Assuming that the step function $K(q/\La)$ (see \eqn{A:props}) varies
very quickly near $q=\La$, we can evaluate it,
\beq
I_N(\La)=\dsp {\La^4\over N} (\La^2 + m^2 + \Si_\La)^{-N}.
\eeq
On the RHS of \Eqn{A:messy}, the quantity in braces contains
$N-1$ summations. If $N=1$ then it evaluates to the quantity
in square brackets, with $l_N=n$. Otherwise the summations
are evaluated in the usual way, with $l_N= n-l_1-\cdots -l_{N-1}$.
Summing over $n$,
\beq
\La {\p V_\La \over \p\La} = \half K_4 \sum_{N=1}^\infty
{-(-1)^N\over N} (\La^2 + m^2 + \Si_\La)^{-N}
\Biggl\{
\sum_{l_1=1}^\infty {\Ga^{2l_1+2}_\La \over (2l_1)!}
\cdots
\sum_{l_N=1}^\infty {\Ga^{2l_N+2}_\La \over (2l_N)!}
\phi^{2(l_1+\cdots +l_N)}
\Biggr\},
\eeq
so by \eqn{A:V},
\beq
\La {\p V_\La(\phi) \over \p\La} =\dsp \half K_4 \La^4 \ln\biggl( 1 +
  {Y''_\La(\phi)\over \La^2 + m^2 + \Si_\La} \biggr).
\eeq
Ignoring terms that are independent of $\phi$,
\beq
\dsp\La {\p V_\La(\phi) \over\p \La} =\dsp \half K_4 \La^4
  \ln\Bigl(\La^2 + V''_\La(\phi)\Bigr),
\eeq
which was to be proved.

\vspace{36pt}

\centerline{\bf Figure Captions}

Figure 1. The general form of the RG flows in the $\{ \la_1, \la_2\}$
subspace are displayed, along with the four fixed points, for the case
$N>4$. Shading indicates the ``unstable'' regions where the quartic
part of the potential is no longer positive definite. The values
$l=l_*$ and $l=l_f$ of the flow parameter are where, respectively,
a stability line is crossed, and $r=\O(1)$ so that the subsequent
flow is canonical ($\{ \la_1(l), \la_2(l)\}$ flow along a straight line
projecting through the origin).

Figure 2. The general form of the RG flows in the
$\{ h, e^2\}$ subspace for the $N$-component
Abelian Higgs model for $N<366$, and in $(4-\ep)$-dimensions.
Both fixed points are unstable. The physical ``unstable'' region
is $h<0$.

\end{document}